\documentclass{article}
\usepackage{spconf,amsmath,epsfig}

\providecommand{\doi}[1]{doi: {\footnotesize \href{http://dx.doi.org/#1}{\path{#1}}}}

\usepackage[pdftex=true,breaklinks=true,hidelinks=true,colorlinks=true,citecolor=blue]{hyperref}

\usepackage[square,numbers]{natbib}

\usepackage[utf8]{inputenc}     

\usepackage{bbold}

\title{QUANTUM MACHINE LEARNING FOR REMOTE SENSING:\\ EXPLORING POTENTIAL AND CHALLENGES}
\name{Artur Miroszewski$^1$, Jakub Nalepa$^{2,3}$, Bertrand Le Saux$^4$, Jakub Mielczarek$^1$ \thanks{This work was funded by the European Space Agency,
and supported by the ESA $\Phi$-lab 
(\url{https://philab.esa.int/}) AI-enhanced 
Quantum Computing for Earth Observation (QC4EO) 
initiative, under ESA contract No. 4000137725/22/NL/GLC/my. 
AM and JM were supported by the Priority Research 
Areas Anthropocene and Digiworld under the program 
Excellence Initiative – Research University at the 
Jagiellonian University in Krak\'ow. JN was supported 
by the Silesian University of Technology grant for 
maintaining and developing research potential.}}
\address{$^1$Institute of Theoretical Physics, Jagiellonian University,
\L{}ojasiewicza 11, 30-348 Cracow, Poland\\
$^2$KP Labs, Bojkowska 37J, 44-100 Gliwice, Poland\\
$^3$Silesian University of Technology, Akademicka 16, 44-100 Gliwice, Poland\\
$^4$European Space Agency $\Phi$-lab, Largo Galileo Galilei 1, 00044 Frascati, Italy}
\begin{document}
%
\maketitle
\begin{abstract}
The industry of quantum technologies is rapidly expanding, offering promising opportunities for various scientific domains. Among these emerging technologies, Quantum Machine Learning (QML) has attracted considerable attention due to its potential to revolutionize data processing and analysis. In this paper, we investigate the application of QML in the field of remote sensing.
It is believed that QML can provide valuable insights for analysis of data from space. We delve into the common beliefs surrounding the quantum advantage in QML for remote sensing and highlight the open challenges that need to be addressed.
To shed light on the challenges, we conduct a study focused on the problem of kernel value concentration, a phenomenon that adversely affects the runtime of quantum computers. Our findings indicate that while this issue negatively impacts quantum computer performance, it does not entirely negate the potential quantum advantage in QML for remote sensing.
\end{abstract}
\section{Introduction}
\label{sec:intro}

Quantum computing technologies attract attention from academia, business entities and general public.
With the constant development of both technology and theory behind quantum computation, the claims for its potential use grow significantly.
At the same time, as those claims get validated, we are becoming more and more aware of the limitations of quantum computing. We review recent developments in the theory of Quantum Machine Learning (QML) and refer them to the subject of remote sensing. We propose the following understanding for quantum advantage in machine learning.
We claim quantum advantage if:\\
    \textbf{1.} the algorithm run on physical quantum machine solves the machine learning task obtaining better performance than on a classic machine,\\
    \textbf{2.} the simulation of the quantum algorithm is not efficient (in terms of cost, energy use, runtime, \dots) on a classic machine.

One of the industries that might profit in employing quantum technologies for pattern recognition and data analysis is remote sensing. With the constantly increasing amount of data produced in space and specific requirements for data handling, it is natural to look for new technologies which could excel in those tasks. Indeed, scientists are already exploring this topic from the perspective of Quantum Machine Learning ~\cite{MissionPlanning, sychang-eo-classif, asebastianelli-eo-classif, Clouds, Gawron, Sozo_1, Sozo_2, Su, Amer_1, Amer_2}. Therefore, we recognize the need for a high-level review of the topic.

In Sec. \ref{sec:QML} we introduce the notion of Quantum Machine Learning and discuss potential advantages of exploring this subject. In Sec. \ref{sec:challenges} we review major challenges for QML to achieve quantum advantage. In Sec. \ref{sec:example} we perform a study on one of those challenges. Sec. \ref{sec:conclusions} concludes the paper.

\section{Quantum Machine Learning}\label{sec:QML}
Quantum machine learning refers to the intersection of quantum computing and machine learning. It encompasses the development and application of machine learning techniques on quantum computers, outsourcing some machine learning tasks to quantum computer to utilize cooperation of quantum and classical machines or using quantum-inspired approaches. The main goal of QML is to explore and exploit the potential advantages offered by quantum computing. The expected source of quantum advantage can be \cite{QML, qalgorithms}:\\
    \textbf{Improved computational efficiency}. Quantum algorithms can potentially solve some computational problems exponentially faster than classic algorithms. This could enable faster training and inference in machine learning tasks, leading to more efficient models and quicker decision-making.\\
    \textbf{Solving complex optimization problems}. Many machine learning tasks involve optimization problems, such as finding the best model parameters or minimizing a cost function. Quantum optimization algorithms, such as the quantum approximate optimization algorithm (QAOA) or the variational quantum eigensolver (VQE), offer the potential for improved solutions or more efficient exploration of the solution space.\\
    \textbf{Enhanced data representation and processing}. Quantum machine learning algorithms can leverage the unique properties of quantum systems, such as quantum superposition and entanglement, to represent and process data in new ways. This could enable more expressive representations, richer feature spaces, and more efficient data processing techniques.\\
    \textbf{Quantum data analysis and pattern recognition}. The amount of accessible quantum states grows exponentially ($2^n$) with the number of qubits $n$ used. 
    It gives an opportunity to densely encode enormous amounts of data into relatively small number of qubits (using, for example, amplitude encoding). 
    Utilizing entanglement can provide a way to represent non-trivial correlations present in the data.
    Additionally, operating on high-dimensional space can provide a way to increase the expressivity of machine learning methods.





\section{Challenges for quantum advantage}\label{sec:challenges} \textbf{Hardware noise}. One of the main reasons for not being able to claim true quantum advantage is the access only to Noisy-Intermediate Scale Quantum hardware. The noise introduced by every circuit operation is still on the high level. There are whole industries devoted to introducing error mitigation and correction to quantum computers. Although excellent development on both hardware and software side of this subject \cite{IBMnature}, we are still far away from obtaining large scale, reliable quantum algorithms on current physical devices.\\
\textbf{Data embedding}. Most of the beautiful, early results showing quantum supremacy \cite{QPCA, QSVMOld} rely on the assumption that the quantum algorithms acts on some readily-prepared quantum states. It is often assumed that some quantum memory or QRAM is used at the start of the quantum routine, yet no such thing exists at the moment. It has been shown \cite{Tang} that this assumption is unrealistic and the state preparation usually introduces polynomial or even exponential overhead, which invalidates the supremacy claims.\\
\textbf{Untrainability and value concentration}. What was initially considered as the biggest quality of quantum machine learning, the expressivity connected to the size of quantum Hilbert space is also the origin of the biggest software problem. It has been found that the training landscape for quantum neural networks \cite{QNN} includes \textit{barren plateaus}~\cite{BarrenPlateaus} as the number of qubits $n$ grow. The optimization function becomes almost everywhere flat, with a deep, yet narrow global minimum. Both gradient-based and gradient-free optimization methods struggle with barren plateaus \cite{gradientfree}, rendering the quantum neural networks untrainable.
For quantum-kernel support vector-based algorithms \cite{havlicek}, it has been found that with the increasing number of qubits $n$ the quantum-kernel non-diagonal entries vanish exponentially fast \cite{concentration}.
The quantum-kernel Gramm matrix $\mathcal{K}$  becomes indistinguishable from the identity matrix $\mathcal{K} \xrightarrow{n\rightarrow \infty} \mathbb{1}$ undermining the machine learning task. The way to keep the kernel matrix elements distinguishable is to exponentially increase the number of measurements performed on a quantum computer, resulting in significant increase in computational time. \\
\textbf{Quantum state measurement}. We do not have direct access to read the quantum state, being the result of computation. To get useful information from it, one has to perform a measurement on the state. 
The measurement collapses the possible superposition state into a single eigenvector of measurement operator. 
This single fact considerably restricts the usefulness of quantum parallelism. 
One can perform quantum computation on many states at the same time, using quantum superposition, but the single measurement gives information only about one state in the mixture.
In many cases, the actual result of computation is the probability of obtaining a specific state. 
Performing the quantum computation and measurement repeatedly allows the estimation of an expectation value of this probability. 
More measurement give better estimate. 
This can pose a serious problem if our algorithm depends on distinguishing the eigenstate probabilities for different experimental setups. If the values of probabilities to distinguish are closer to each other, we need more measurements to differentiate between them.
The shot noise in quantum measurement originates from the Born rule. Therefore, it is an inherent property of quantum computing and cannot be eliminated by hardware improvements of quantum computers. \\
\textbf{Performance of classical algorithms}. Artificial intelligence is constantly developing, but in its core it is a mature technology, extremely well optimized for classical machines. Trying to compete with efficient classical algorithms is hard, especially that quantum computing is still in its infancy, both on the software and hardware side. So far, there has been no serious claim about quantum advantage in the real-life generic industry problem.
In remote sensing, the quantum algorithms have been tested in cloud \cite{Clouds} and land cover \cite{Gawron, Su, Amer_1, Sozo_1} classification tasks, satellite mission planning \cite{MissionPlanning}, and many other tasks.
Yet, the results show, that the quantum machine learning algorithms can compete with their classical counterparts, or even show better performance on artificially created datasets, the true advantage for generic problems and datasets was not found.

\section{Example: Quantum advantage for quantum-kernel SVM}\label{sec:example}

\begin{figure*}[t]
\begin{minipage}[b]{.45\linewidth}
  \centering
  \includegraphics[width=0.9\columnwidth]{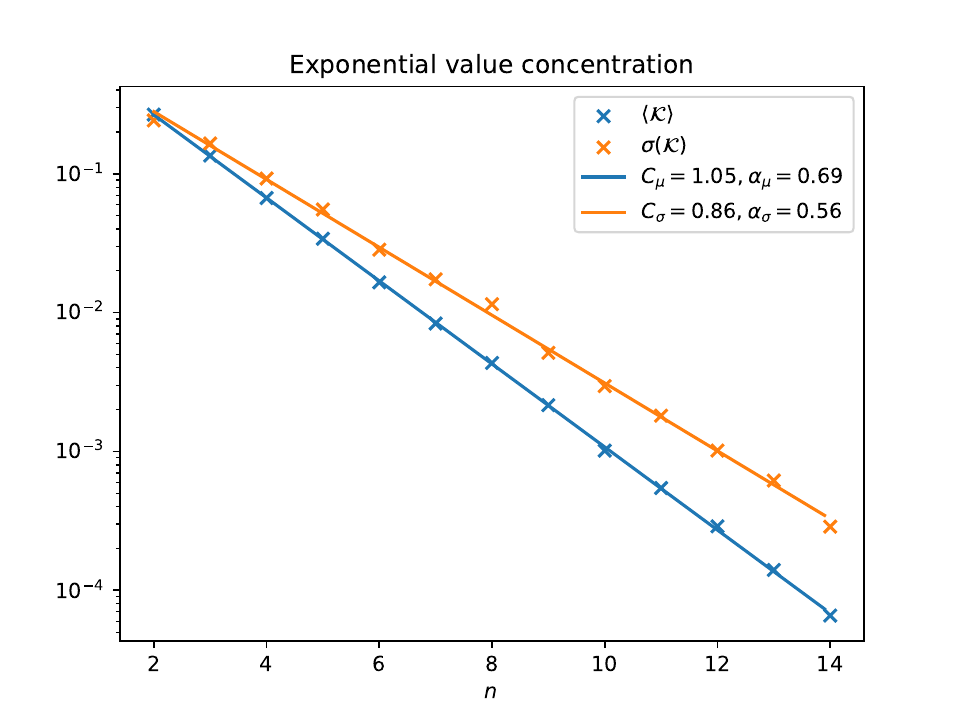}
 \label{res1}
  \centerline{(a)}\medskip
\end{minipage}
\hfill
\begin{minipage}[b]{0.45\linewidth}
  \centering
  \includegraphics[width=0.9\columnwidth]{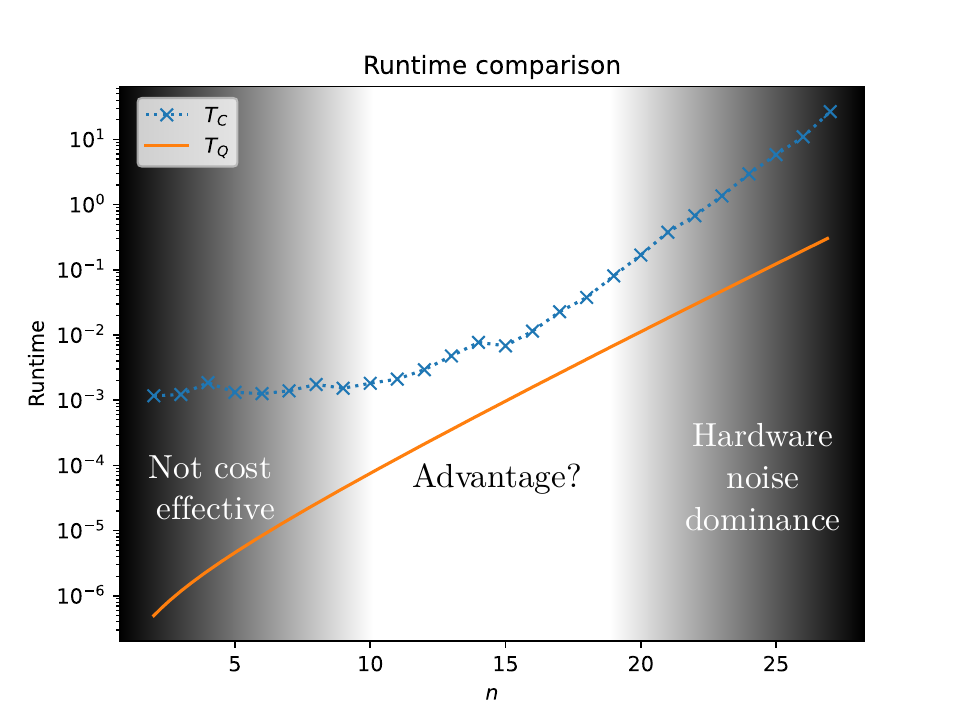}
  \centerline{(b)}\medskip
\end{minipage}
\caption{Experimental results for quantum-kernel SVM. (a) The mean value $\langle \mathcal{K} \rangle$ and standard deviation $\sigma(\mathcal{K})$ of independent kernel entries vanish exponentially with the number of qubits $n$ used. The fit $C_X e^{-\alpha_X n}$ for both of the scalings is presented. (b) Comparison of the numerically obtained classical simulation time with the estimated quantum runtime (Eq. \ref{eq:TQ}). Even in the light of recent "no-go" results \cite{concentration}, the simple model gives hope for a possible quantum advantage.}
\label{fig:res}
\end{figure*}
For the sake of the example, we propose the following, rather speculative, assumption. In some classification task, we were able to obtain better accuracy using SVMs with quantum kernels than with any other state-of-the-art classical method.
In the view of kernel value concentration problem, we would like to assess whether it is faster to estimate the kernels on a quantum computer, or simulate them on a classical machine as the number of the used qubits ($n$) grows. We consider fidelity kernels $\mathcal{K}(x,x')=|\langle \phi(x')|\phi(x)\rangle|^2$ induced by a ZZ-feature map, which is proven to be $\#P$-hard to simulate \cite{havlicek}. We show the kernel value concentration on $m=100$ artificially generated data points. The classical computation time $T_C(n)$ is obtained by numerical experiments performed on a desktop computer with \texttt{Qiskit} package.

We propose a simple model for estimating quantum computing time $T_Q(n)$ of a single kernel entry. It consists of single circuit runtime $t_{circ}(n)$ multiplied by a number of times the circuit was run $R(n)$:
\begin{equation}\label{eq:TQ}
    T_Q(n) = R(n)\cdot t_{circ}(n) = R(n)\left[N_l(n)\cdot t_g+t_m \right],
\end{equation}
where $t_g$ is the time of execution of single gate/layer in the circuit, $t_m$ is a final measurement duration, and 
\begin{equation}
N_l(n) = 4+6(n-1)
\end{equation}
\noindent is the number of layers in the kernel-estimating circuit with the linearly entangled ZZ-feature map.

In Fig. \ref{fig:res}(a), we can appreciate that indeed both the mean value of independent kernel entries $\langle \mathcal{K} \rangle = \frac{2}{m(m-1)}\sum_{i>j} \mathcal{K}_{ij}$ and their standard deviation $\sigma(\mathcal{K})$ vanish exponentially. Hence, in order to keep independent kernel entries distinguishable, we have to increase a precision at which we estimate them on a quantum computer. 

Treating the estimated kernel value $\Tilde{\mathcal{K}}_{ij}$ as a random variable in the Bernoulli process with $R(n)$ experiment repetitions, we know that 
\begin{equation}
\Tilde{\mathcal{K}}_{ij} \xrightarrow{R\rightarrow \infty} \mathcal{K}_{ij},\ \sigma(\Tilde{\mathcal{K}}_{ij}) = \sqrt{\frac{\mathcal{K}_{ij}(1-\mathcal{K}_{ij})}{R(n)}}.
\end{equation}
Now, we have two scales that we can connect. If \textit{precision ratio} $\gamma \doteq \frac{\sigma(\mathcal{K})}{\sigma(\Tilde{\mathcal{K}}_{ij})} \gg 1$ then the spread in the kernel values is greater than the uncertainty in measurement of their individual values. Therefore they are distinguishable.

If we model the exponential vanishing of the mean of independent entries and their standard deviation as $C_{\mu} e^{-\alpha_{\mu} n}$ and $C_{\sigma} e^{-\alpha_{\sigma} n}$ respectively we can obtain an expression for the needed number of circuit runs for the given precision of estimating kernel entry
\begin{equation}
    R(n) = \gamma^2 \frac{C_{\mu}}{C_{\sigma}^2}e^{(2\alpha_{\sigma}-\alpha_{\mu})n}\left[1-C_{\mu}e^{-\alpha_{\mu}n}\right].
\end{equation}
Substituting this formula to Eq. \ref{eq:TQ} we can estimate quantum computing runtime.
The exponential vanishing coefficients are obtained from the kernel simulations in Fig. \ref{fig:res}(a), while gate execution and measurement times are estimated as $t_g = 10^{-8}s, t_m = 10^{-7}s$ \cite{times}, the \textit{precision ratio} is set to $\gamma=10$.

In Fig. \ref{fig:res}(b), we show the comparison between classical simulation and estimated quantum computation runtime. Both of the runtimes start to grow exponentially around number of qubits $n=15$, but the crucial observation is that, even with the exponential concentration of kernel entries, the quantum runtime scaling is better than classical one.

We confirmed the existence of the exponential concentration of independent kernel values in kernels obtained with quantum circuits. We introduced a simple model for runtime estimation for a quantum computer. The model takes into account the need for increasing the number of experiment repetitions stemming from the exponential value concentration. Using the current literature and our own simulations, we were able to find the values of free parameters in the model and draw a curve of quantum runtime $T_Q$ as a function of the size of quantum machine. We compared $T_Q$ with numerically obtained time of classical simulation $T_C$. Both $T_Q$ and $T_C$ scale exponentially, therefore there is no hope for quantum supremacy, where the classic and quantum algorithms are in different complexity classes. Nevertheless, the growth rate for classic simulation is greater, giving hope for quantum advantage with decent speedups. Although $T_Q$ in Fig. \ref{fig:res}(b) is always smaller than $T_C$ we acknowledge that our simple model does not take into account the cost of quantum computation nor the effect of hardware noise. Therefore, we expect that for small $n$ performing quantum computation would not be cost effective, while for big $n$ the hardware noise will hinder the possible advantage. We postulate that in the intermediate size of quantum machines there is a room for quantum advantage.

\section{Conclusions}\label{sec:conclusions}
We outlined the opportunities and challenges of QML for its potential use in remote sensing applications.
One has to be careful in the claims of quantum advantage, as under throughout investigation, many of original beliefs are weakened by current results in the theory of quantum computation.
On the other hand, as we show in the case of kernel value concentration problem, the weakening of the quantum advantage does not mean its complete vanishing. Therefore, we stay optimistic about the applications of QML in real-life applications.


\bibliographystyle{plainnat}
\bibliography{myBibFile}

\end{document}